\documentclass[prb,floatfix,superscriptaddress,citeautoscript,twocolumn]{revtex4-1}
\usepackage{amssymb,amsmath,amsthm,amsfonts}
\usepackage{graphicx}
\usepackage{graphics}
\usepackage{color}
\usepackage{caption}
\usepackage{verbatim}
\usepackage[lofdepth,lotdepth,caption=false]{subfig}
\usepackage[pdfstartview=FitH]{hyperref}
\usepackage[usenames,dvipsnames,svgnames,table]{xcolor}
\usepackage{placeins}
\usepackage{comment}
\usepackage{txfonts} 
\usepackage[cal=cm]{mathalfa} 
\usepackage{cleveref} 
\crefname{equation}{Eq.}{Eqs.}
\crefname{figure}{Fig.}{Figs.}
\usepackage{hyperref}

\newsavebox{\bigleftbox}
\hypersetup{
 colorlinks,
 citecolor=Blue,
 linkcolor=Blue,
 urlcolor=Blue
}

\begin{document}

\title{Tuning Penta-Graphene Electronic Properties Through Engineered Line Defects}

\author{Ramiro Marcelo dos Santos}
\affiliation{Institute of Physics, University of Bras\'{i}lia , 70.919-970, Bras\'{i}lia, Brazil}
\author{Leonardo Evaristo de Sousa}
\affiliation{Theoretical and Structural Chemistry Group, State University of Goias, 75133-050 Anapolis, Brazil}
\author{Douglas Soares Galv\~{a}o}
\affiliation{Applied Physics Department, State University of Campinas, Campinas, SP, 13083-959, Brazil}
\author{Luiz Antonio Ribeiro Junior}
\affiliation{Institute of Physics, University of Bras\'{i}lia , 70.919-970, Bras\'{i}lia, Brazil}

\date{\today}

\begin{abstract}
Penta-graphene is a quasi-two-dimensional carbon allotrope consisting of a pentagonal lattice in which both $sp^2$ and $sp^3$-like carbons are present. Unlike graphene, penta-graphene exhibits a non-zero bandgap, which opens the possibility of its use in optoelectronic applications. However, as the observed bandgap is large, gap tuning strategies such as doping are required. In this work, density functional theory calculations are used to determine the effects of the different number of line defects of substitutional nitrogen or silicon atoms on the penta-graphene electronic behavior. Our results show that this doping can induce semiconductor, semimetallic, or metallic behavior depending on the doping atom and targeted hybridization ($sp^2$ or $sp^3$-like carbons). In particular, we observed that nitrogen doping of $sp^2$-like carbons atoms can produce a bandgap modulation between semimetallic and semiconductor behavior. These results show that engineering line defects can be an effective way to tune penta-graphene electronic behavior.
\end{abstract}

\maketitle

\section{Introduction}

Research on 2D materials has gained much attention since the discovery of graphene\cite{novoselov_Science,geim_NM}. This carbon allotrope that is composed of a hexagonal lattice was demonstrated to possess several exciting features, such as high electrical and thermal conductivity and large mechanical resistance that are aimed at developing the next generation of organic optoelectronic devices \cite{neto2009electronic,balandin2008superior}. However, when it comes to optoelectronic applications, graphene has a zero bandgap, which precludes its use, for instance, as active material in solar cells. To overcome this issue, gap opening strategies have been developed \cite{tuning_gap_graphene1, tuning_gap_graphene2, tuning_gap_graphene3}, including the cutting of graphene sheets into graphene nanoribbons \cite{son2006energy} and the use of dopants \cite{DENIS2010251,HOUMAD20151021,doi:10.1063/1.4791011}. 

\begin{figure*}[!htb]
\centering
\includegraphics[width=1.0\linewidth]{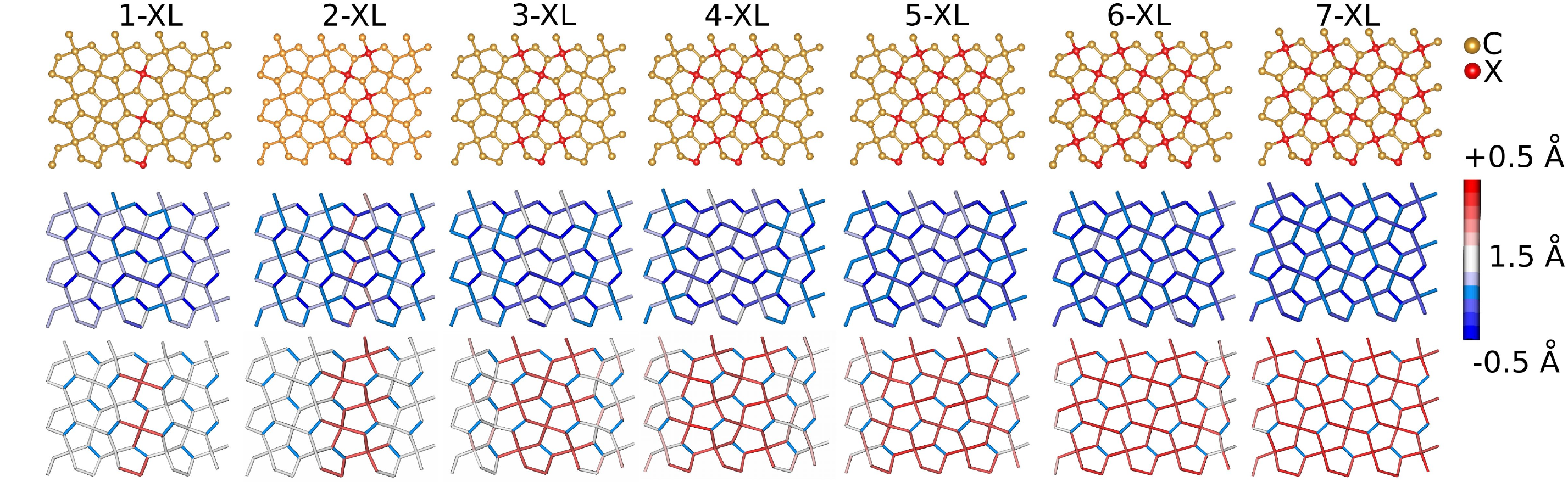}
\caption{(Top) Schematic representation for the $sp^3$-like carbons doping strategy. The red color represents the doping atoms (N or Si). Bond length variations for penta-graphene lattices with nitrogen and silicon doping schemes are presented in the middle and bottom panels, respectively. In the color palette on the right, the equilibrium distance is 1.575 \AA. The maximum deviations are 0.5 \AA~for nitrogen and 1.5 \AA~for silicon doping schemes.}
\label{morfsp3}
\end{figure*}

There are a few nature-occurring carbon allotropes, including graphite, fullerenes \cite{kroto1985c60}, and carbon nanotubes \cite{iijima1993single}. Recently, another allotrope was proposed named penta-graphene \cite{pentagraphene} (see Figure \ref{morfsp3}), which has a lattice composed of pentagons that resemble the Cairo pentagonal tiling. It is supposed to present ultrahigh ideal strength even above that of graphene and to be able to withstand temperatures of up to 1000 K \cite{1}. Penta-graphene was not sensitized yet, and it is believed that it can be at least a metastable material \cite{1,cranford20166,rajbanshi2016energetic,ewels2015predicting}. Furthermore, recent studies have suggested that chemical functionalization employing hydrogenation or fluorination could even increase its structural stability\cite{einollahzadeh2016studying,2}. Electronic structure calculations indicate that penta-graphene is a semiconductor presenting bandgap values about  2.24--4.3 eV \cite{1,rajbanshi2016energetic,einollahzadeh2016studying}. Such large bandgap suggests the need for gap tuning strategies to be developed in order to tailor this material for optoelectronic applications. In this sense, the particular topology of penta-graphene, which presents both $sp^2$ and $sp^3$-like carbon hybridizations in its lattice, raises the question of how its structural and electronic properties will behave by targeting a particular hybridization in the doping processes.

Herein, we carried out density functional theory (DFT) calculations to address the effects of the systematic substitutional doping of either $sp^2$ or $sp^3$-like carbons by nitrogen and silicon atoms in a penta-graphene lattice. Mainly, we investigate the effect of the different number of engineered line defects on their electronic and structural properties (see Figure \ref{morfsp3}). Our findings show that, in terms of morphology, nitrogen doping is responsible for increasing the stiffness of the lattice in comparison to pristine penta-graphene, whereas silicon doping results in the simultaneous stretching and compression of Si--C and C--C bonds, respectively, concerning undoped C--C bonds. The two doping schemes investigated here (Si or N) produce significantly different results in terms of electronic behavior. Silicon doping allows us to tune the bandgap when replacing $sp^3$-like carbons and produces metallic behavior when replacing $sp^2$-like ones.  Nitrogen doping replacing $sp^3$-like carbons results in a transition from semiconductor to semimetallic to a metallic character. Nitrogen doping replacing $sp^2$-like carbons produces an alternating behavior between semimetallic and semiconductor depending on the number of dopants. These results indicate that engineered line defects can be a very effective way to tune penta-graphene electronic behavior.

\section{Computational Details}
DFT calculations were carried out within the Generalized Gradient Approximation (GGA) scheme as proposed for Perdew, Burke, and Ernzerhof (GGA/PBE) \cite{perdew96} along with the DZP \cite{hohenberg64,Kohn65} basis set. Relativistic pseudopotentials parameterized within the Troullier-Martins formalism were also considered \cite{kleinman82}. These combined approximations can accurately describe the magnetic and electronic properties of materials composed of atoms with many electrons. All the calculations were performed considering spin polarization. For the bands and density of states calculations, Monkhorst pack grid of 21 x 21 x 3 was used \cite{monkhorst76}. A mesh cutoff of 200 Ry was chosen as a parameter for our calculations \cite{anglada02}. The force criteria convergence was 0.001 eV/ \AA. In order to establish a good compromise between the accuracy of our results and computational costs, the tolerance in the matrix density and total energy was set to 0.0001 and 0.00001 eV, respectively. All calculations were performed with the SIESTA software suite \cite{ordejon96,portal97}.

\section{Results and Discussion}
In the present work, we investigated the electronic and structural features of penta-graphene lattices with substitutional doping (N or Si) either at $sp^3$ or $sp^2$-like carbons forming engineered line defects. The cases of 1 up to 7 line defects were considered. In this sense, Figure \ref{morfsp3} (top panels) shows a schematic representation of these defective structures for the $sp^3$ case, which is similar to the $sp^2$ one. Figure \ref{morfsp3} presents seven scenarios, which are identified by N--XL, with N corresponding the number of line defects in the horizontal/vertical directions, and X refers to either nitrogen or silicon dopant atoms.  

Significant structural differences take place in the morphology of the resulting doped structures. For the nitrogen cases, an overall decrease in bond length values was observed, as presented in Figure \ref{morfsp3} (middle panels), where the blue color represents bond length compression when compared to the original non-doped ones. As expected, these effects are more pronounced around the defect lines, but they extend to the other bonds as the number of doping atoms increases. Such a reduction in bond length values can reach 0.5 \AA, which corresponds to a roughly 30\% decrease concerning the 1.575 \AA~ equilibrium distance found in pristine penta-graphene. In contrast, for the silicon atoms, the newly formed C-Si bonds undergo a substantial expansion of up to 1.5 \AA, which would amount to 100\% increase in bond length values when compared to pristine penta-graphene C-C bonds, as shown in Figure \ref{morfsp3}-bottom. For the remaining C-C bonds, this effect is less pronounced.

\begin{figure*}[!htb]
\centering
\includegraphics[width=0.7\linewidth]{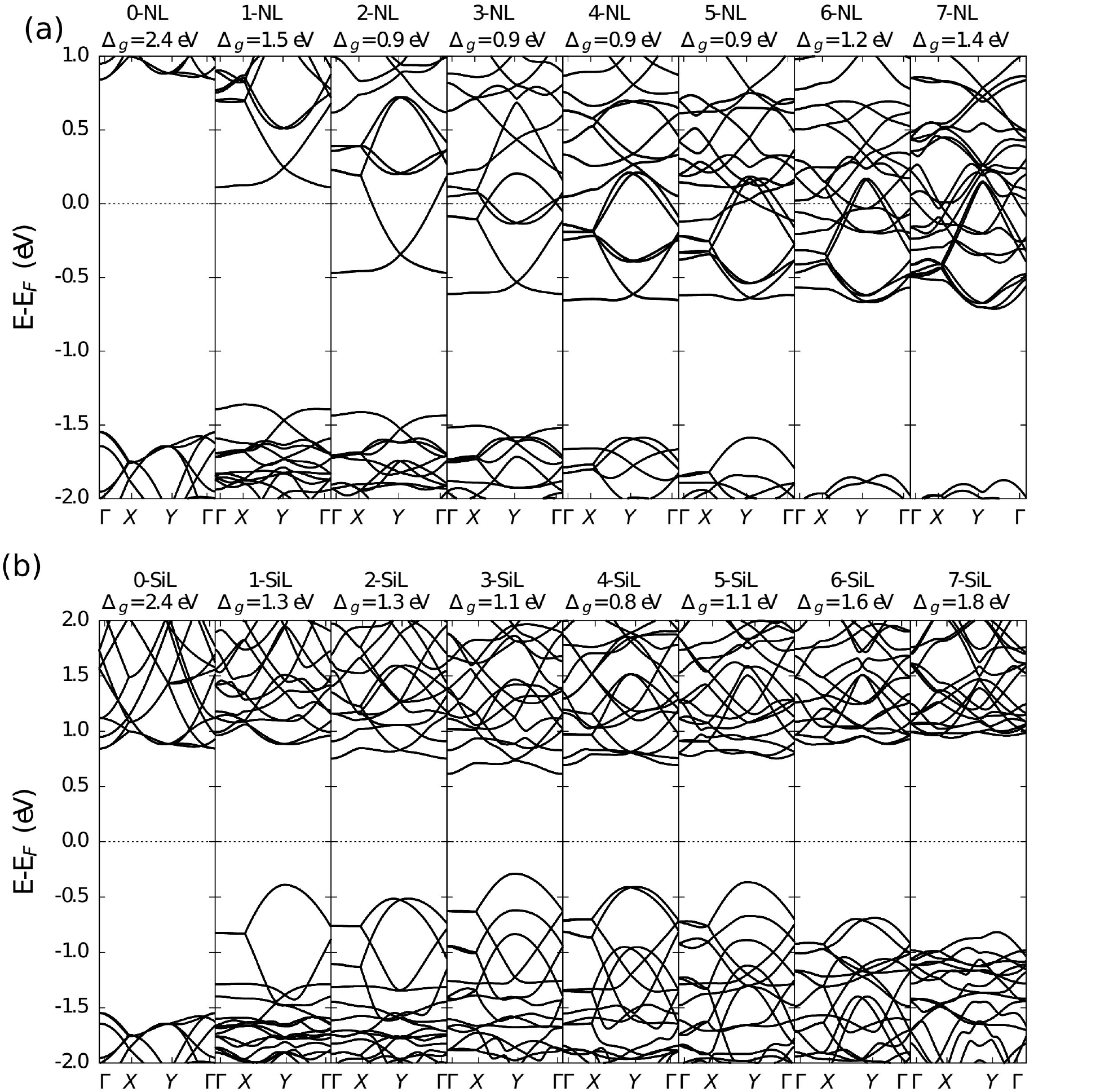}
\caption{\label{bandasp3} Band structures for undoped and doped penta-graphene lattices, as a function of the number of line defects for the $sp^3$ case. (a) and (b) correspond to the results for nitrogen and silicon doping schemes, respectively. Here, X-NL and X-SiL denote the number (X) of dopant lines systematically inserted into the penta-graphene structure.}
\end{figure*}

It is well-known that structural modifications lead to changes in electronic properties. As such, useful information about these changes can be gained by contrasting the band structures of the doped and undoped penta-graphene sheets.  Reports in the literature for penta-graphene indicate a semiconductor material with a bandgap of about 2.4 eV \cite{pentagraphene,rajbanshi2016energetic}.  We have obtained a similar value, as shown in Figure \ref{bandasp3}. In this figure, we also present the band results for the $sp^3$ cases for nitrogen (Figure \ref{bandasp3}-a) and silicon doping (Figure \ref{bandasp3}-b) as a function of the number of line defects. As we can see from Figure \ref{bandasp3}, for the nitrogen with 1 line defect, the bandgap becomes indirect and decreases from 2.4 eV to just 1.5 eV. The Fermi level lies near the conduction band, making this doped penta-graphene structure an n-type semiconductor. As the number of line defects increases, the material no longer presents a bandgap. We can see that for 2 and 3 defect lines, the partial density of states (PDOS) near the Fermi level, presented in Figure S1a in the Electronic Supporting Information (SI), almost disappears. As the number of defect lines increases, the PDOS around the Fermi levels increases, and the doped penta-graphene becomes fully metallic.

The silicon doping of $sp^3$-like carbons, on the other hand, does not result in metallic materials. As seen in Figure \ref{bandasp3}-b and Figure S2a of the SI, doped penta-graphene lattices preserve their semiconductor nature. For the 1 line defect case, the only observed effect was the decrease in bandgap to 1.3 eV, with further doping making the bandgap indirect. Interestingly, the bandgap dependence on the number of defect lines has a convex nature, with a minimum bandgap of 0.8 eV observed for the case of 4 defect lines. After this point, the bandgap increases again, reaching 1.76 eV for the lattice with seven defect lines. The decreases in bandgap values are associated with the appearance of states in the valence band closer to the Fermi level.  

\begin{figure*}[!htb]
\centering
\includegraphics[width=1.0\linewidth]{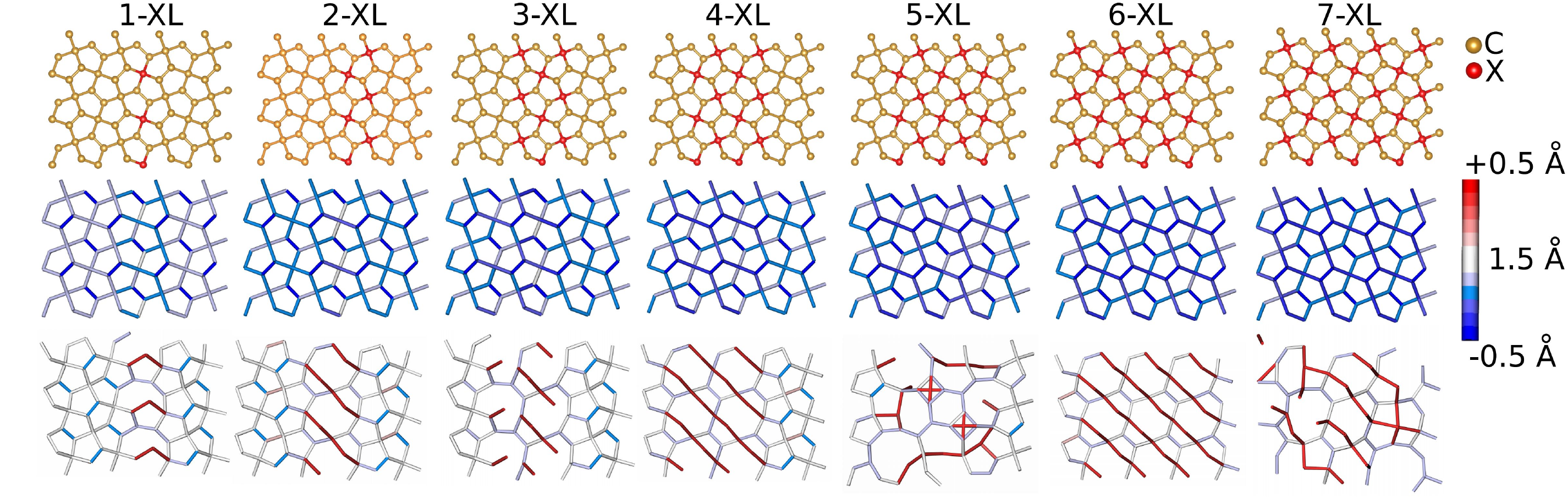}
\caption{(Top) Schematic representation for penta-graphene lattices with $sp^2$-like carbons doping strategy. The red color represents the doping atoms (N or Si). Bond length variations for penta-graphene lattices with nitrogen and silicon doping schemes are presented in the middle and bottom panels, respectively. In the color palette on the right, the equilibrium distance is 1.575 \AA. The maximum deviations are 0.5 \AA~for nitrogen and 1.5 \AA~for silicon doping schemes.}
\label{sp2}
\end{figure*}

\begin{figure*}[!htb]
\centering
\includegraphics[width=0.7\linewidth]{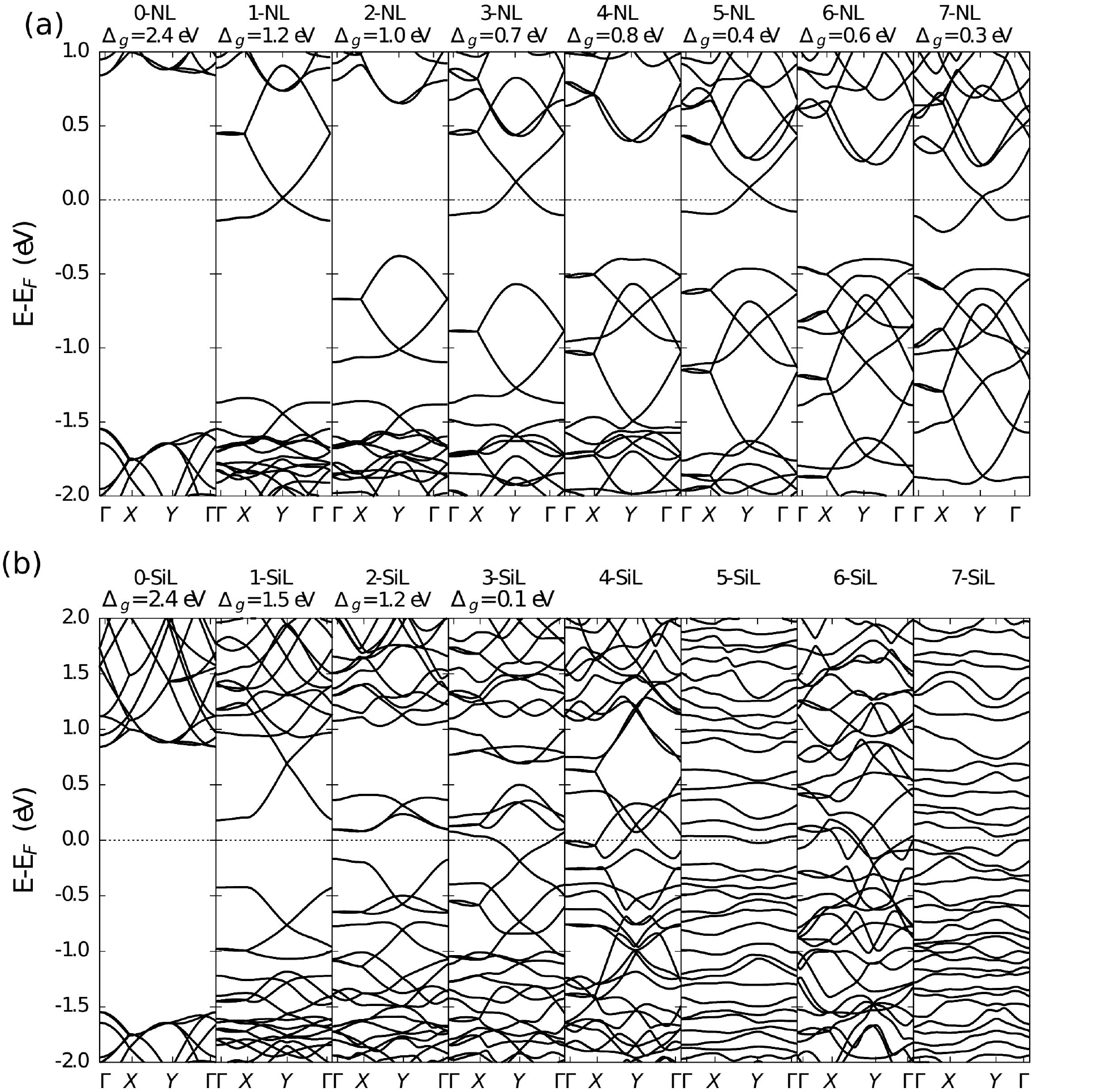}
\caption{Band structures for undoped and doped penta-graphene, as a function of the number of line defects for the $sp^2$ case. (a) and (b) correspond to the results for nitrogen and silicon doping schemes, respectively. Here, X-NL and X-SiL denote the number (X) of dopant lines systematically inserted into the penta-graphene structure.}
\label{bandassp2}
\end{figure*}

The second doping strategy considered here consists of the doping of $sp^2$-like carbons. One main difference is the possibility of having N--N and Si--Si bonds, not present for $sp^3$ case. The seven analyzed scenarios are presented in Figure \ref{sp2} (top panels). Again, nitrogen doping results in overall contraction of the bond length values up to 0.5 \AA~concerning undoped penta-graphene ones. These deviations in the bond lengths are represented in Figure \ref{sp2} (middle panels) by the blue bonds. Silicon doping, in contrast, produces different patterns. These are mostly characterized by the simultaneous expansion of Si--C bonds, by Si--Si bonds that preserve the original 1.575 \AA~bond length of pristine penta-graphene and by C-C bonds that slightly contract. The combination of these effects results in the pattern characterized by blue hexagons diagonally sliced by red Si--C bonds, as depicted in Figure \ref{sp2}-bottom, especially for the six defect lines case. However, two particular cases, 5 and 7 defect lines, break the pattern producing much more disordered configurations likely induced by symmetry breaking. The Si--C distances in the doped penta-graphene lattices can reach 3.8 \AA, considerably larger than Si--C bonds found, for instance, in disilicon carbide, which can be as large as 2.2 \AA\cite{koput2017ab}. This is suggestive that the atoms are no longer bonded and the structures undergo structural rearrangements.

In terms of electronic structure, nitrogen and silicon doping produce completely different results. Interestingly, for the $sp^2$ nitrogen doping cases, the bandgap values exhibit a bandgap modulation (alternating increasing/decreasing) behavior, as can be seen in Figure \ref{bandassp2}-a. For even values of $N$, semiconducting properties are obtained with almost direct bandgaps that decrease as $N$ grows larger from 2.4 eV for $N=0$ to 0.6 eV for $N=6$. In contrast, for odd values of $N$, the valence band maximum (VBM) touches or surpasses the Fermi levels. However, the DOS near the Fermi level is very small, increasing progressively with $N$. This behavior can be better visualized in the PDOS plots of Figure S1b in the SI. The even $N$ nitrogen-doped penta-graphene lattices display a semimetallic character, with carbon $p$ orbitals being mostly responsible for the DOS in the vicinity of the Fermi level. 

Silicon doped penta-graphene structures, on the other hand, possess semiconductor properties only for $N=1$ or $N=2$, with a fast transition to bandgap closing. For larger $N$ values, the structures become fully metallic, as shown in Figure \ref{bandassp2}-b (see also Figure2-ESI in SI). It is also worth noting that the effect of the disordered morphology observed for the 5-SiL and 7-SiL cases is to make energy levels practically independent of momentum when compared to the other cases, as evidenced by the almost flat (dispersionless) bands. In this sense, silicon doping of $sp^2$-like carbons is not as effective as in cases of nitrogen doping regarding decreasing the bandgap value. 

\begin{figure}[!htb]
\centering
\includegraphics[width=0.48\textwidth]{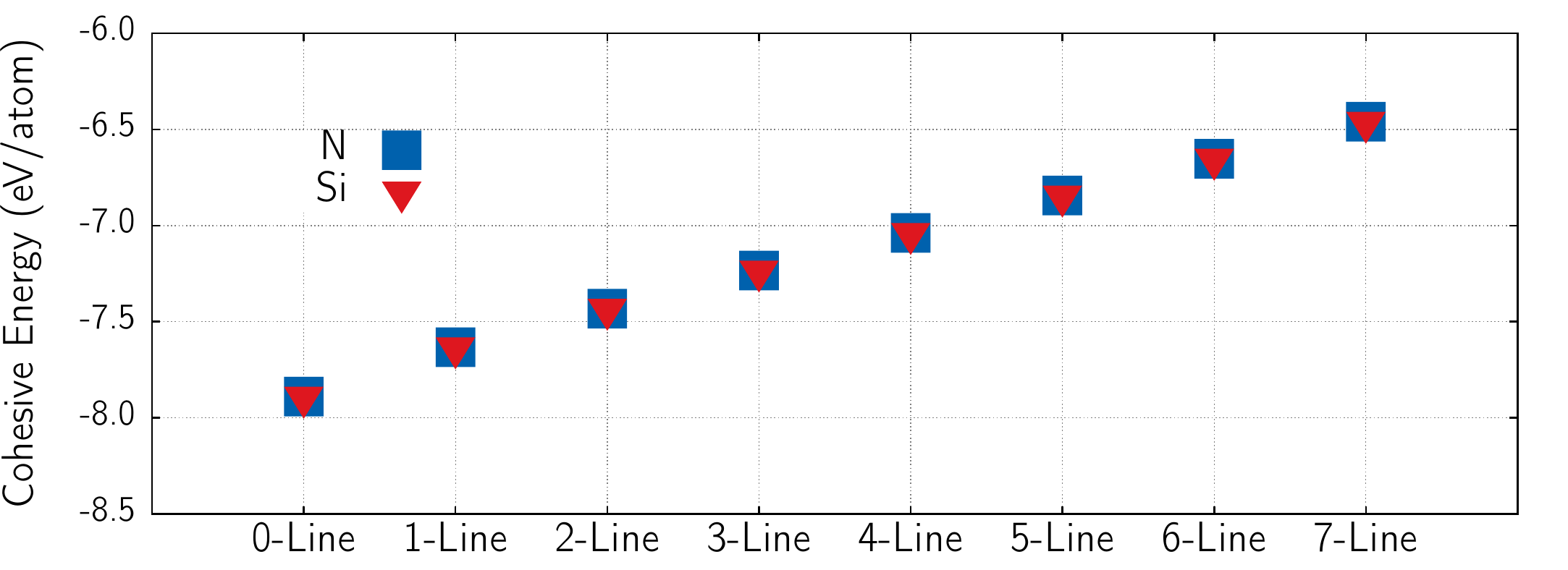}
\includegraphics[width=0.48\textwidth]{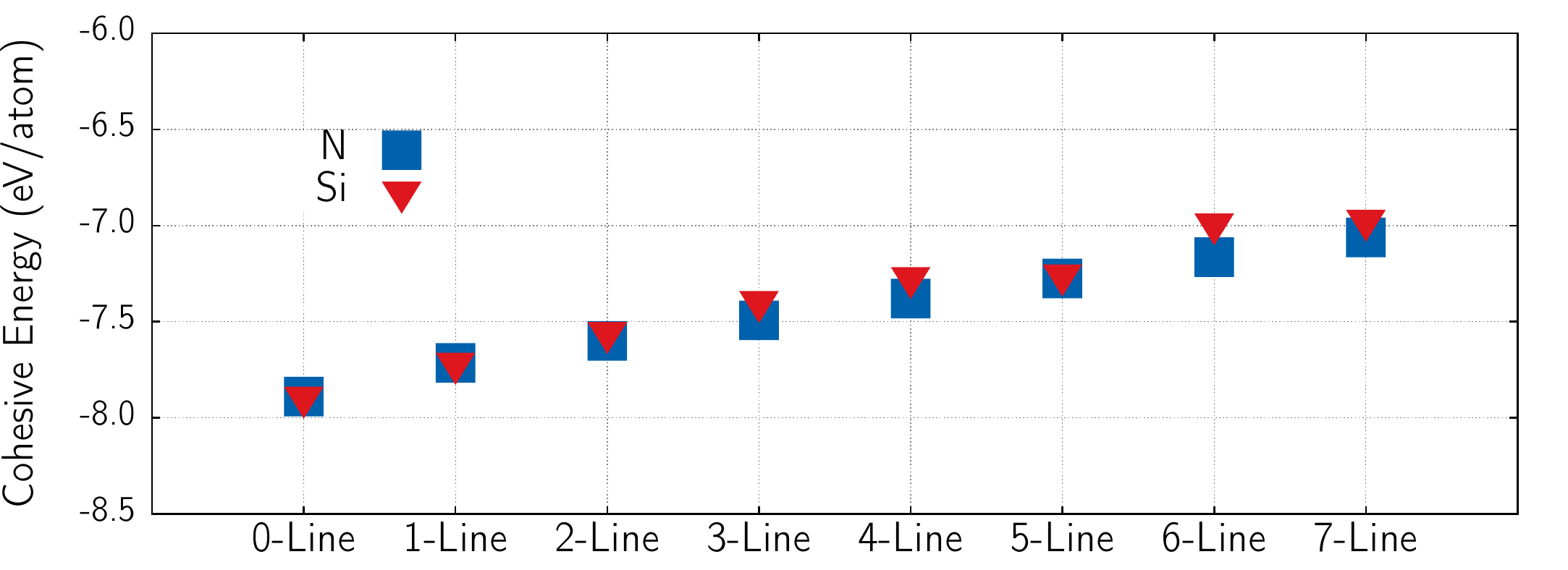}
\caption{Cohesive energy for all cases of nitrogen and silicon doping schemes of (top) $sp^3$ carbons and (bottom) $sp^2$-like carbons.}
\label{spec1}
\end{figure}

Finally, an important advantage concerning the $sp^2$ doping instead of $sp^3$ one is the effects on the structural stability of the doped structures. In Figure \ref{spec1}, we present the cohesive energy values for all doping scenarios. The cohesive energy is the difference per atom between the entire system energy and the sum of the individual energies of its constituents. Figure \ref{spec1}-top) and Figure \ref{spec1}-bottom) refer to $sp^3$ and $sp^2$ results, respectively. It can be seen from this figure that for both $sp^3$ and $sp^2$-like carbon doping, the effects on cohesive energy are similar for N and Si dopants. However, whereas $sp^3$-like carbon doping increases the cohesive energy from -8.0 eV/atom to -6.5 eV/atom, the corresponding increase in the case of $sp^2$-like carbon in less than 1.0 eV/atom.      

\section{Summary and Conclusions}
In summary, we have carried out DFT calculations to assess the changes in the structural and electronic properties of penta-graphene lattices resulting from selective N and Si doping (engineered line defects) of either $sp^3$ or $sp^2$-like carbons. Regardless of the type doping, for the nitrogen cases, we observed an overall stiffening of the penta-graphene structures. On the other hand, for the silicon doping cases, we observed only the stretching of Si-C bonds and compression of the remaining C-C bonds.

From an electronic structure perspective, both the doping type and doping atom selection produce significantly different results. Silicon doping of the $sp^3$-like carbons preserves the penta-graphene semiconductor character. The smallest observed bandgap value was 0.8 eV (2.4 eV for pristine penta-graphene) for the 4 line defect case. For a larger number of line defects, the bandgaps increase again. On the other hand, for Si doping $sp^2$-like carbons, the result is a fast transition to metallic behavior. 

Nitrogen doping produces more interesting results. For the $sp^3$-like carbons, the doping results in a progressive changes from semiconductor to metallic behavior. For the case of 1 line defects, the doped penta-graphene structure becomes an n-type semiconductor with a 1.5 eV bandgap. For 2 and 3 line defects, the bandgap closes, but the density of states near the Fermi level goes to zero, giving the material a semi-metallic character. From this point on, further doping leads to true metallic behavior.

For the nitrogen doping of $sp^2$-like carbons, we observed a bandgap modulation behavior (alternating increase/decrease). For even numbers of line defects, the bandgap decreases with the number of line defects. In contrast, for odd numbers of line defects, a semi-metallic behavior is observed, combining zero bandgap with near-zero density of states at the vicinity of the Fermi level. Finally, the cohesive energy values indicate that doping the $sp^2$-like carbon affects less the structural stability of the resulting doped structures than $sp^3$-like carbon ones.

These results indicate that selective doping of penta-graphene structures through engineered line defects can be an effective tool to tune their electronic behavior, being possible to create structures that vary from large bandgaps through semiconductors and even metallic or semimetallic ones. We hope the present work can stimulate further studies along these lines. 
 
\section*{Conflicts of interest}
There are no conflicts to declare.

\section*{Acknowledgments}
The authors gratefully acknowledge the financial support from Brazilian Research Councils CNPq, CAPES, and FAP-DF. L.A.R.J. gratefully acknowledges the financial support from FAP-DF grant 00193.0000248/2019-32 and the financial support from CNPq grant 302236/2018-0.DSG thank the Center for Computational Engineering and Sciences at Unicamp for financial support through the FAPESP/CEPID Grant \#2013/08293-7.

\bibliography{references}

\newpage

\begin{figure*}[!htb]
\centering
\includegraphics[width=1\linewidth]{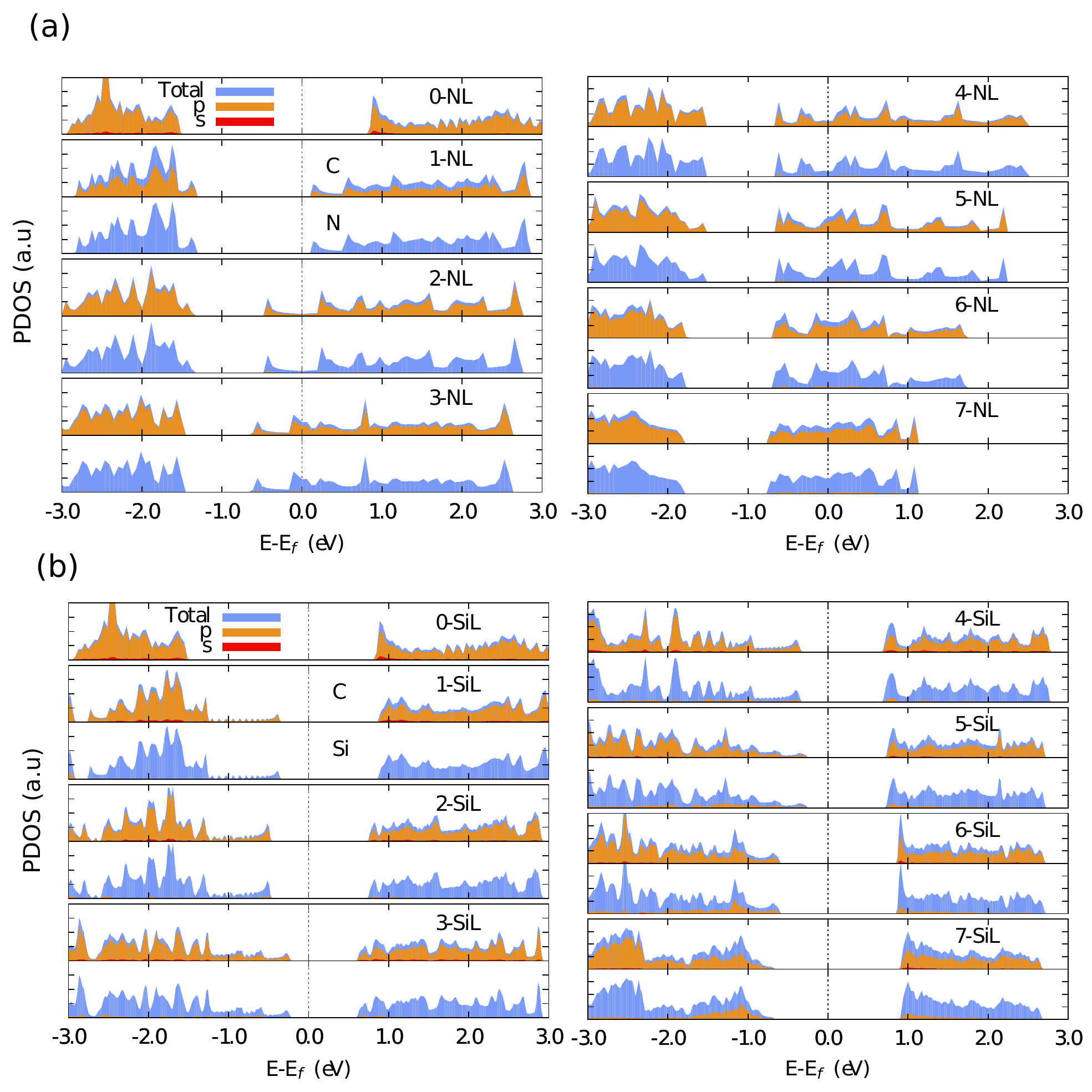}
\captionsetup{labelformat=empty}
\caption{Electronic Supporting Information: Projected Density of States (PDOS) for all schemes of $sp^3$ carbon doping considering both (a) Nitrogen and (b) Silicon atoms. In this figure, X-NL and X-SiL denote the number (X) of dopant lines systemttcally inserted into the pentra-graphene structure.}
\end{figure*}

\begin{figure*}[!htb]
\centering
\includegraphics[width=1\linewidth]{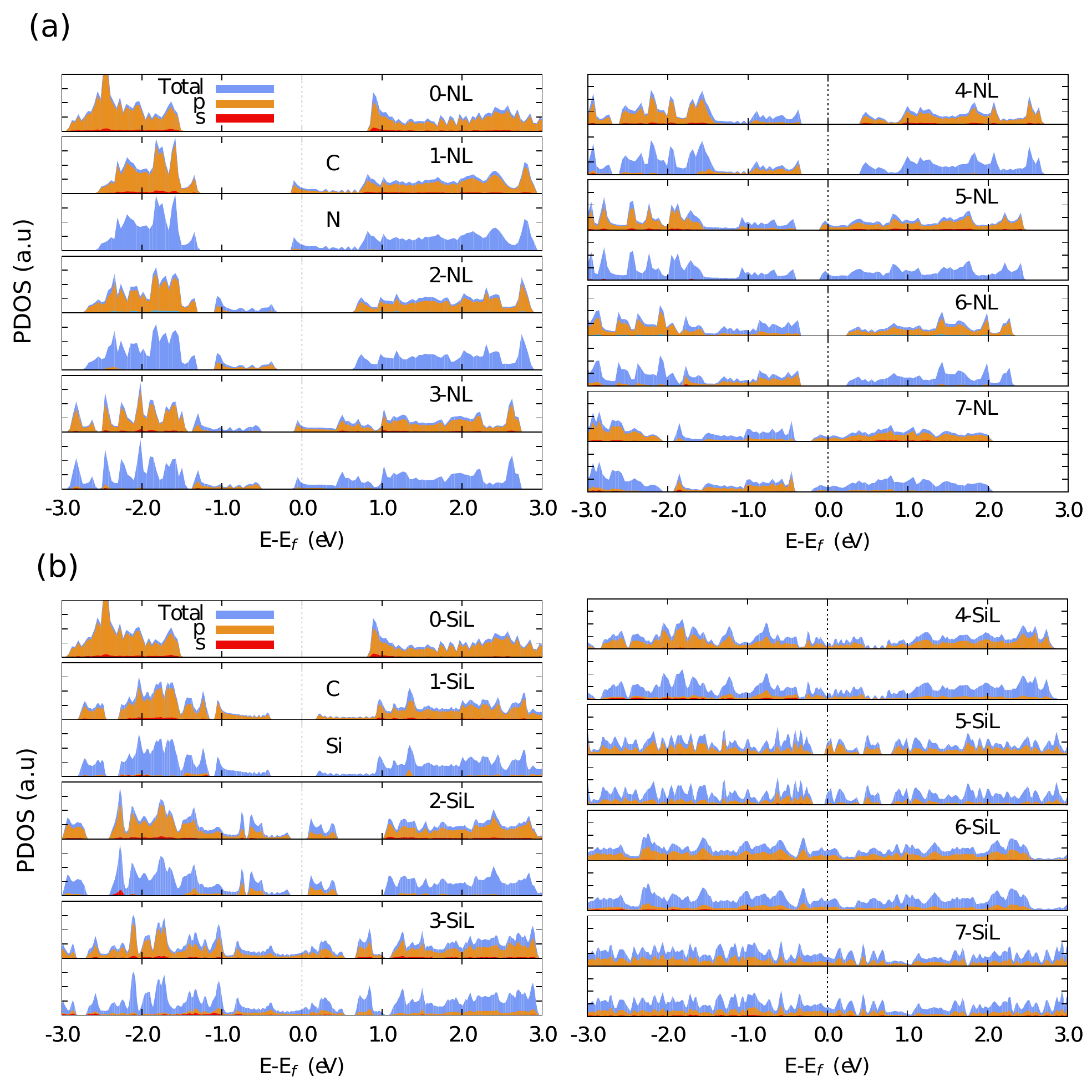}
\captionsetup{labelformat=empty}
\caption{Electronic Supporting Information: Projected Density of States (PDOS) for all schemes of $sp^2$ carbon doping considering both (a) Nitrogen and (b) Silicon atoms. In this figure, X-NL and X-SiL denote the number (X) of dopant lines systemttcally inserted into the pentra-graphene structure. }
\end{figure*}

\end{document}